\documentclass[%
 reprint,
 amsmath,amssymb,
 aps,
prb,
]{revtex4-1}

\usepackage{amsmath}
\usepackage{graphicx}
\usepackage{dcolumn}
\usepackage{bm}
\usepackage{natbib}
\usepackage{url}
\usepackage[colorlinks,citecolor=red,linkcolor=blue,urlcolor=blue]{hyperref}
\usepackage{caption}
\usepackage{hhline}
\usepackage{float}
\usepackage{mhchem}
\usepackage{calrsfs}
	\DeclareMathAlphabet{\pazocal}{OMS}{zplm}{m}{n}
\usepackage{physics}
\usepackage{appendix}
\usepackage[export]{adjustbox}
\newcommand{\cdag}[1]{\hat{c}^\dagger_{#1,s}}
\newcommand{\cndag}[1]{\hat{c}_{#1,s}}

\newcommand{\dXdY}[2]{\frac{\partial #1}{\partial #2}}

\newcommand{\squeeze}[3]{\left\langle #1\middle| #2\middle| #3\right\rangle}
\newcommand{\uu}[1]{u^{#1}_{s}(\bold{k})}

\newcommand{\cdagU}[1]{\hat{c}^\dagger_{#1,\uparrow}}
\newcommand{\cdagD}[1]{\hat{c}^\dagger_{#1,\downarrow}}
\newcommand{\cndagU}[1]{\hat{c}_{#1,\uparrow}}
\newcommand{\cndagD}[1]{\hat{c}_{#1,\downarrow}}
\newcommand{\CHI}[0]{\hat{\chi}_{ij}}
\newcommand{\CHIU}[0]{\hat{\chi}_{ij,\uparrow}}
\newcommand{\CHID}[0]{\hat{\chi}_{ij,\downarrow}}

\begin{document}


\title{Anti-chiral edge states in Heisenberg ferromagnet on a honeycomb lattice}

\author{Dhiman Bhowmick}
\author{Pinaki Sengupta}%
\affiliation{%
 School of Physical and Mathematical Sciences, Nanyang Technological University, Singapore \\
}%

\date{\today}

\begin{abstract}
We demonstrate the emergence of anti-chiral edge states in
a Heisenberg ferromagnet with  
Dzyaloshinskii–Moriya interaction~(DMI) on a honeycomb lattice
with inequivalent sublattices. The DMI, which acts 
between atoms of the same species, differs in magnitude for the two 
sublattices, resulting in a shifting of the energy of the magnon bands 
in opposite directions at the two Dirac points. The chiral symmetry
is broken and for sufficiently strong asymmetry, the band shifting
leads to anti-chiral edge states (in addition to the normal chiral 
edge states) in a rectangular strip where the magnon current propagates 
in the same direction 
along the two edges. This is compensated by a counter-propagating 
bulk current that is enabled by the broken chiral symmetry. We analyze
the resulting magnon current profile across the width of the system
in details and suggest realistic experimental probes to detect them. 
Finally, we discuss about possible materials that can potentially exhibit such 
anti-chiral edge states. 
\begin{description}
\item[PACS numbers]
 85.75.-d, 75.47.-m, 73.43.-f, 72.20.-i
\end{description}
\end{abstract}

\pacs{Valid PACS appear here}
\maketitle


\paragraph*{\label{sec1}Introduction.-}
Quantum magnets have emerged as a versatile platform for realizing 
magnetic analogues of the plethora of topological phases that have been
predicted, analysed, classified and observed in electronic systems over
the past decade. Haldane's paradigmatic model\cite{Haldane} of tight binding electrons
on a honeycomb lattice with complex next-nearest neighbor hopping -- that
constitutes the foundation of many of the electronic topological phases --
has a natural realization in (quasi-) 2D insulating ferromagnets such as 
\ce{CrI3}\cite{CrI3} and \ce{AFe2(PO4)2}  (A=Ba,Cs,K,La)\cite{IronDirac}. 
In many of these materials, the dominant
Heisenberg exchange is supplemented by a next-nearest
neighbour anti-symmetric DMI. 
Magnetic excitations in these systems are described by two species 
of quasi-particles -- spinons with up and down spins. The 
Kane-Mele-Haldane model -- analogous to the Kane Mele model for 
electrons -- has been proposed to describe the spinons over a wide 
range of temperatures.\cite{Kane_Mele_Haldane} The spinon bands
acquire a non-trivial dispersion due to Berry phase arising from the 
DMI. This results in a spin Nernst effect (SNE) where a thermal 
gradient drives a transverse spin current, a spinon version of the 
spin Hall effect\cite{Kane_Mele_Haldane,SNE1,SNE2,SNE3}. 
In a finite sample, the two spinon species generate two 
counterpropagating spin currents along the edges that are 
protected by chiral symmetry of the Hamiltonian -- analogous to two copies
of the thermal Hall effect (THE) of magnons that has 
been observed in many insulating magnets\cite{THE1,THE2,THE3,THE4}.  .

Recently, there has been growing interest in engineering systems with
co-propagating edge currents\cite{Colomes,AntiChiral1,AntiChiral2}, through
an ingenious, yet physically unrealistic, modification of the Haldane
model. The conservation of net current is 
satisfied by counter propagating bulk current. That is, the bulk
is not insulating, in contrast to conventional topological insulators.   
In this work, we demonstrate that anti-chiral states arise naturally
in spinons on a honeycomb magnet comprised of two different magnetic ions,
with unequal DMI for the two sublattices. In the absence of DMI, the spinon dispersion
consists of two doubly degenerate bands with linear band crossings at
{\bf K} and {\bf K$^\prime$}\cite{DiracMagnon}. A finite DMI lifts the 
degeneracy between the two spinon branches and opens up a gap in the
spectrum\cite{opendirac1,opendirac2,opendirac3,opendirac4}. For 
{\it asymmetric} DMI, the two bands
for each spinon species are shifted in opposite directions relative
to each other at the {\bf K} and {\bf K$^\prime$} points in the Brillouin zone.
This results in similar dispersion for the gapless modes at both edges,
giving rise to co-propagating edge states.
This is shown to yield effective anti-chiral edge states
for the spinons in addition to normal chiral ones. We present a detailed
characterization of the nature of the edge and bulk spinon states and 
suggest suitable experimental signatures to detect these novel topological
states.

\begin{figure}[H]
	\centering
	\includegraphics[width=0.35\textwidth]{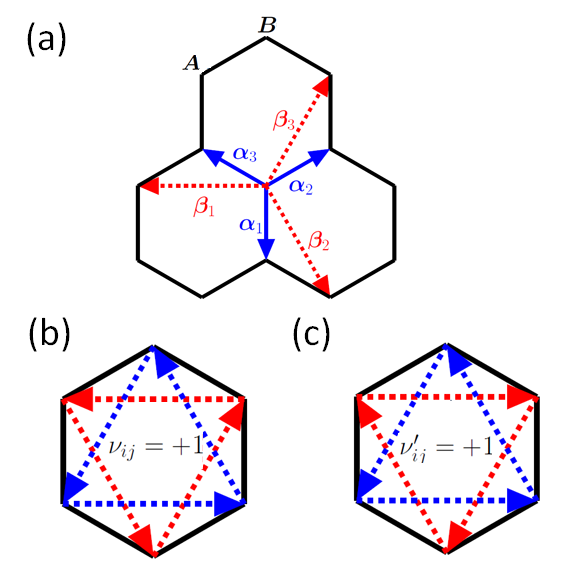}
	\caption{(color online)(a) The honeycomb lattice structure. (b) The directions along which $\nu_{ij}=+1$ has been shown, otherwise $\nu_{ij}=-1$. (c) The directions along which $\nu_{ij}'=+1$ has been shown, otherwise $\nu_{ij}'=-1$.}
	\label{lattice}
\end{figure}

\paragraph*{\label{sec2}Model.-}
We consider a Heisenberg ferromagnet on the honeycomb lattice with
unequal DMI ($D_A$ and $D_B$) on the two sub-lattices. 
Introducing the symmetric and anti-symmetric combinations of $D_A$ and $D_B$ as,
$D~=~\frac{1}{2}\left( D_A+D_B \right)$, $D'~=~\frac{1}{2} \left( D_A-D_B \right)$ -- 
termed chiral and anti-chiral DMI respectively for reasons that will become
clear later -- the Hamiltonian is given by,

\begin{align}
\pazocal{H}=& -J\sum_{\left\langle i,j \right\rangle} \bold{S}_i\cdot\bold{S}_j+D\sum_{\left\langle\left\langle i,j \right\rangle\right\rangle} \nu_{ij} \hat{z}\cdot \left(\bold{S}_i\times \bold{S}_j\right) \nonumber \\
& +D'\sum_{\left\langle\left\langle i,j \right\rangle\right\rangle} \nu_{ij}' \hat{z}\cdot \left(\bold{S}_i\times \bold{S}_j\right)-B\sum_i S_i^z,
\label{eq:hamil}
\end{align}
where, $J>0$ is the nearest neighbor Heisenberg interaction and $\nu_{ij}=+1$ when $i$ and $j$ are along the cyclic arrows shown in Fig.\ref{lattice}(b). Finally, $\nu_{ij}'=+\nu_{ij}$ for sublattice-A and $\nu_{ij}'=-\nu_{ij}$ for sublattice-B. The magnetic field $B$ is introduced in a Zeeman coupling term to stabilize the ferromagnetic ground state at finite temperature. The energy scale is set
by choosing $J=1$ --  all other parameters in the Hamiltonian are in units of $J$. 

The ground state of the hamiltonian (Eq.\ref{eq:hamil}) is ferromagnetic for
$J > -{3\sqrt{3} \over 2}\sum_s \left | D + sD'\right |,\quad s=\pm 1$.
We apply the Schwinger Boson mean field theory~(SBMFT) to study the topological
character of the low energy magnetic excitations at a finite temperature.
The Schwinger Boson representation consists of the mapping the spin 
operators into spinons as,
$
S_i^+ =c_{i,\uparrow}^\dagger c_{i,\downarrow}$,
$S_i^- =c_{i,\downarrow}^\dagger c_{i,\uparrow}$,
$S_i^z =\frac{1}{2}\left(c_{i,\uparrow}^\dagger c_{i,\uparrow}-c_{i,\downarrow}^\dagger c_{i,\downarrow}\right),
$
where $c_{i,s}$ and $c_{i,s}^\dagger$ are the annihilation and creation operators 
of spin-1/2 up or down spinons respectively. The constraint 
$\sum_{s} c_{i,s}^\dagger c_{i,s}=2S$, $\forall i$ on the bosonic operators 
ensures the fulfillment of the spin-S algebra.

 After applying Schwinger Boson transformation along with the constraint, and
 using a mean field approximation to reduce the 4-body operators to bilinear forms,
 the spin model Eq.\ref{eq:hamil} is mapped to the the mean field hamiltonian,

\begin{align}
\pazocal{H}=&-\eta J\sum_{\left\langle i,j \right\rangle,s} \left[ \cdag{i} \cndag{j}+\mathtt{H.c.} \right] +\sum_{i,s} \left(\lambda-\frac{sB}{2}\right) \cdag{i}\cndag{i}\nonumber\\
&+\frac{D}{2}\sum_{\left\langle\left\langle ij \right\rangle\right\rangle, s}\left[ \left(i \nu_{ij} s \zeta_{-s} + s\xi_{-s} \right)  \cdag{i}\cndag{j}+\mathtt{H.c.}\right]\nonumber \\
&+\frac{D'}{2}\sum_{\left\langle\left\langle ij \right\rangle\right\rangle, s}\left[ \left( i \nu'_{ij} s \zeta_{-s} + s\xi'_{-s} \right) \cdag{i}\cndag{j}+\mathtt{H.c.}\right]
\label{eq:spinonH}
\end{align}
where the mean field parameters are defined as, $\eta=\sum_s\left\langle \hat{\chi}_{ij,s} \right\rangle\equiv\sum_s \left\langle \cdag{i} \cndag{j}\right\rangle$ evaluated 
on the nearest neighbour-bonds,  and $\zeta=\zeta'=\frac{1}{2}\left\langle \hat{\chi}_{ij,s}+\hat{\chi}_{ji,s} \right\rangle$,  
$\xi=\frac{\nu_{ij}}{2i}\left\langle \hat{\chi}_{ij,s}-\hat{\chi}_{ji,s} \right\rangle$ 
and $\xi'=\frac{\nu'_{ij}}{2i}\left\langle \hat{\chi}_{ij,s}-\hat{\chi}_{ji,s} \right\rangle$, 
evaluated on next nearest neighbour bonds. The terms associated with the parmeters $\eta, \nu_{ij}\zeta_{-s} $ of spinon Hamiltonian
Eq\ref{eq:spinonH} constitute the Kane-Mele-Haldane model\cite{Kane_Mele_Haldane}. 
The term with parameter $\nu'_{ij}\zeta_{-s}$ corresponds to the anti-chiral hopping term introduced in Ref.\onlinecite{Colomes}. The terms with the parameters $\xi_s$ and $\xi'_s$ have no effect on the energy or 
the topological character of the bands, as the parameters are found to be much
smaller compared to other mean field parameters.    
$\lambda$ is the Lagrange undetermined multiplier introduced to implement the 
local constraint. The mean field parameters are obtained by solving a set of self-consistent equations, derived by minimizing the Helmholtz free energy at a particular temperature\cite{SM}.

\begin{figure}[H]
	\centering
	\includegraphics[width=0.5\textwidth]{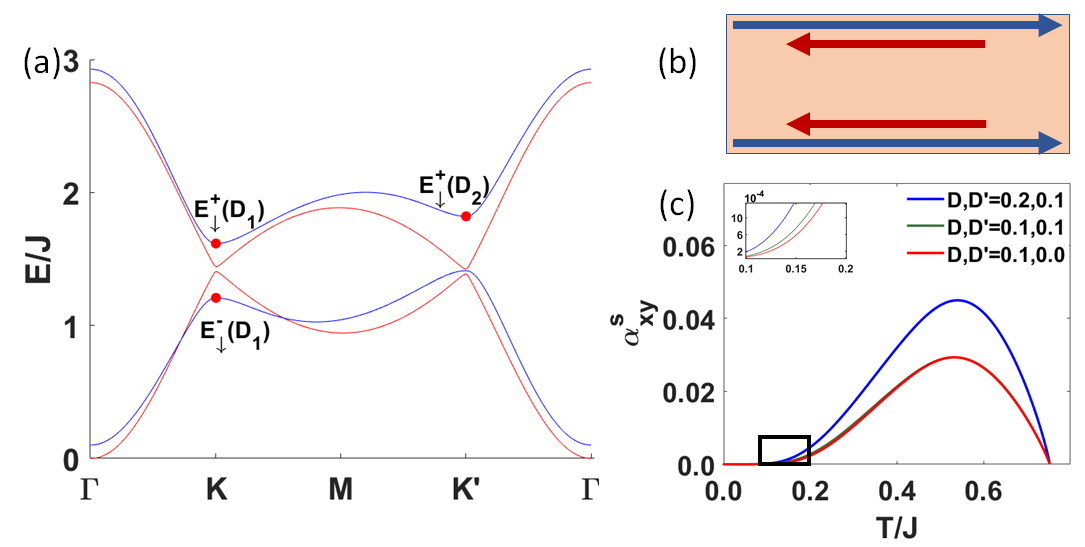}
	\caption{(color online)(a)band along symmetry lines 
	$\Gamma K$, $KM$,$MK'$,$K'\Gamma$ for $J=1.0, B=0.1, D_{ch}=0.1,D_{Ach}=0.05, T=0.25$. The blue band is for down-spinon band and the red band is for up-spinon band. (b) The schematic of the antichiral contribution of the current from edge-states(blue-arrows) and bulk-states(red-arrows). (c) Nernst-conductivity v.s. temperature plot for different DMIs. The inset shows magnified figure of the rectangular portion of the figure.}
	\label{band}
\end{figure}

\paragraph*{\label{sec4} Results.-}
Band structure for spinons at a temperature T=0.25J is shown in  Fig.\ref{band}(a).
In the absence of DMI, the two bands
cross linearly at the Dirac points {\bf K} and {\bf K$^\prime$}\cite{DiracMagnon}. A finite DMI opens up a gap with magnitude 
$\Delta_{s}=3\sqrt{3}\left|D\zeta_{-s}\right|$ in each spinon sector at {\bf K} 
and {\bf K$^\prime$}\cite{opendirac1,opendirac2,opendirac3,opendirac4,Kane_Mele_Haldane}. 
For {\it anisotropic} system ($D_A\neq D_B$) considered here, the gap opening
is not symmetric and leads to a {\it tilting} of the spinon bands near the Dirac 
momenta. The band tilting for each 
band in each spinon sector, defined as the energy difference between 
two Dirac-points in the same band, is given by, $T_{s}^\tau=3\sqrt{3}\left|D'\zeta_{-s}\right|$. While the anti-chiral 
DMI drives the tilting of the bands, it has no effect on 
the magnitude of the band gap. 
Crucially, the tilting is opposite for the two species of spinons. 
For the parameters chosen in Fig.\ref{band}(a), the gap and tilting for 
the up-spinon bands are smaller than those for the 
down spinon bands. This is because in the presence of positive magnetic 
field $B=0.1$ considered here, there are
fewer down spinons and consequently, $\zeta_{\downarrow}<\zeta_{\uparrow}$.

The bands in each spinon sector carry non-zero Berry curvature.
Spin Nernst effect has been proposed as a 
physical phenomenon to identify Berry curvature of spinon bands 
when there are comparable numbers of up 
and down spinons. Here we explore whether it can detect the existence of
anti-chiral DMI. The Nernst conductivity has been calculated using the  
expression $\mathbin{\alpha^s_{xy}}{=\frac{1}{2V}\sum_{\bold{k},s,\tau}s c_1[\rho^\tau_s(\bold{k})]\Omega^\tau_s(\bold{k})}$\cite{Kane_Mele_Haldane}$^,$\cite{Kovalev}, where $\mathbin{\rho^\tau_s(\bold{k})}{=1/(\exp(E^\tau_s(\bold{k})/T)-1)}$ is the Bose-Einstein distribution and $\mathbin{\Omega^\tau_s(\bold{k})}{=i\sum_{\tau' \neq \tau}\frac{\squeeze{\uu{\tau}}{\dXdY{\pazocal{H}}{k_x}}{\uu{\tau'}}\squeeze{\uu{\tau'}}{\dXdY{\pazocal{H}}{k_y}}{\uu{\tau}}-(k_x\leftrightarrow k_y)}{(E^\tau_s(\bold{k})-E^{\tau'}_s(\bold{k}))^2}}$. The results are plotted in Fig.\ref{band}(c) for different $D$ and $D'$. Increase in $D$ increases the band gap as well as the Berry curvature away from the Dirac-points. 
As a result, the Nernst conductivity is substantially affected by $D$ (Fig.\ref{band}(c)). Conversely, since the Berry curvature is independent of 
$D'$, the anti-chiral DMI has very little effect in the Nernst conductivity. The effect of $D'$ on Nernst conductivity can be observed at low temperature due to tilting of the band structure(inset of Fig.\ref{band}(c)). But at higher temperature, the $D'$ has no influence in Nernst conductivity, because the contributions from higher bands overshadows the effects of band tilting.  So, the presence of $D'$ in the system is very hard to detect using Nernst conductivity. Instead, we suggest an alternative way to detect the presence of antichiral DMI.

The gapped bands are topologically non-trivial with Chern numbers $C_{\uparrow}^-=+1,C_{\uparrow}^+=-1, C_{\downarrow}^-=-1, C_{\downarrow}^+=+1$\cite{hatsugai}. 
Due to bulk-edge correspondence, we expect to observe edge states in a 
finite system. In the isotropic limit ($D_A=D_B$),
the edge states are topologically protected by a chiral symmetry. The spinon currents
along the two edges are equal and opposite for the up and down 
spinons. This results in a net flow of spins along the two edges in 
opposite directions -- any scattering to the bulk states is prevented
by symmetry constraints. For the asymmetric system considered here,
$D'$ induces an anti-chiral edge current of spinons where each
species of spinon flows in the same direction along the two edges.
This is balanced by couterflow current of spinons in the opposite direction
carried by the bulk modes. The anti-chiral DMI breaks the chiral symmetry 
protecting the edge states and enables scattering between edge and bulk states. 
This edge-to-bulk scattering produces the bulk current that balances the
anti-chiral edge current.
In the following we discuss how the bulk and edge state dispersion changes
due to interplay between the chiral and anti-chiral DMI.

\begin{figure}[H]
	\includegraphics[width=0.5\textwidth , left]{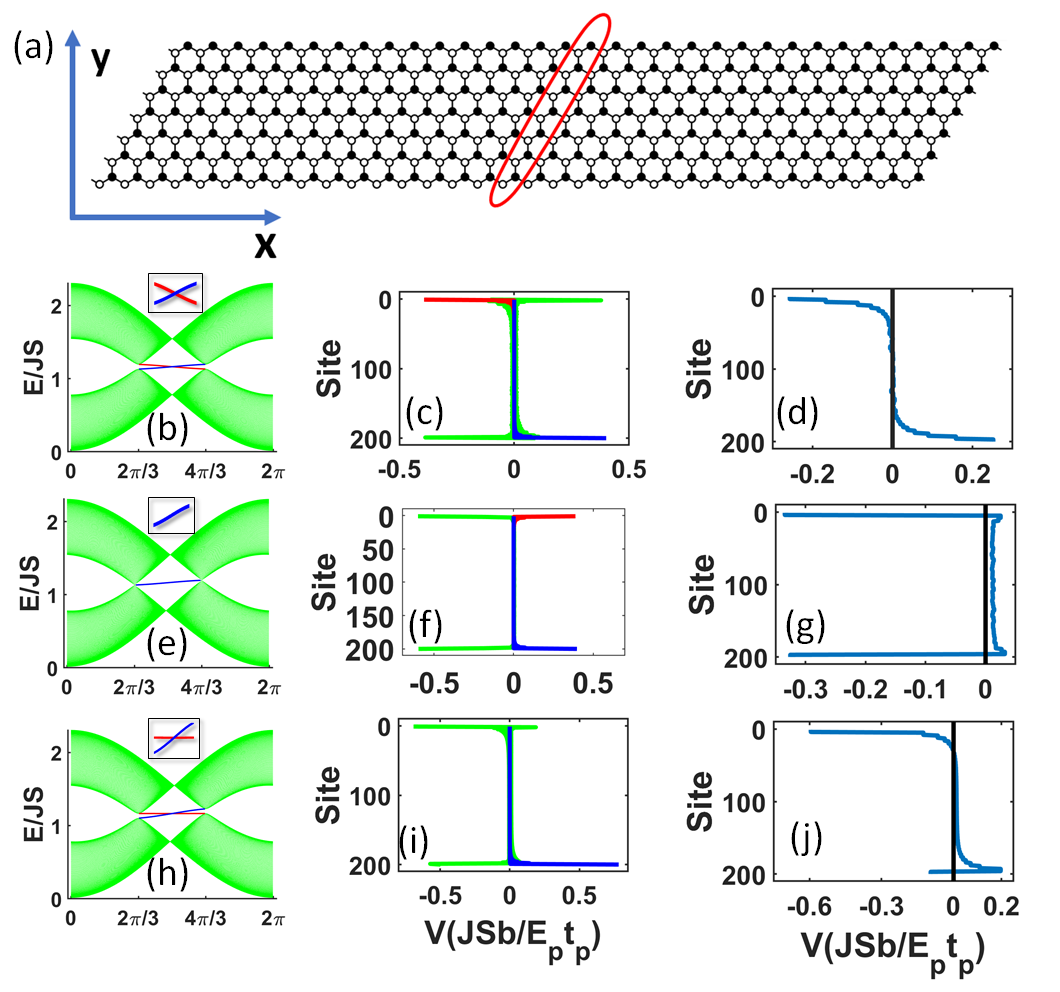}
	\caption{(color online)(a) A honeycomb ribbon. The encircled sites are the basis of unit cell. The figure sets $\left\lbrace(b)-(d)\right\rbrace$,$\left\lbrace(e)-(g)\right\rbrace$,$\left\lbrace(h)-(j)\right\rbrace$ represents result for down-spinon from $200\times 500$ stripe with DMIs $\left\lbrace D=0.1,D'=0.001\right\rbrace$,$\left\lbrace D=0.001,D'=0.1\right\rbrace$,
	$\left\lbrace D=0.1,D'=0.1\right\rbrace$ respectively. The other parameters are $J=1.0,  B=0.1, T=0.5$ for all the plots. The figure sets $\left\lbrace(b),(e),(h)\right\rbrace$ shows the band structure and inset of the figures shows the magnified dispersion of the edge states. The figure sets $\left\lbrace(c),(f),(i)\right\rbrace$ gives the spatial current distribution along width of the stripe. The figure sets $\left\lbrace(d),(g),(j)\right\rbrace$ shows the average of spatial current distribution over four sites respectively. In the figure sets $\left\lbrace(b),(c),(e),(f),(h),(i)\right\rbrace$ the green, red and blue plots corresponds to bulk-state, upper edge edge-state and lower-edge edge-state respectively. Moreover, the results for up-spinon is qualitatively same, the only difference is the dispersions are opposite for the bulk and edge states.}
	\label{edge_state}
\end{figure}

Fig.~\ref{edge_state} shows the spinon bands for a honeycomb nano-ribbon 
with dimension $200\times 500$ lattice sites with zigzag edges, together
with the spin current profile
along the width of the ribbon. Three different sets of $(D, D')$
are chosen to illustrate the evolution of band dispersion and spin
currents with changing DMI. For clarity of presentation, only one
species of spinons is illustrated. Along with the total spin current, the 
contributions from the bulk and two edge modes are calculated separately to 
identify the effects of $D'$ on each component. The spinon
bands and the individual spin currents are color coded for easy 
identification. Green represents the bulk bands and their
contribution to the spin current at each position along the width of the 
ribbon; red (blue) denotes the localized spinon mode and the associated
spin current at the top (bottom) edge. A negative (positive) value 
of the spin current denotes spinon transport to the left (right) along
the length of the ribbon. 

For $D>D'$, the tilting of the bands is
small and the dispersion of edge states at 
upper and lower edges are opposite, as shown in the  Fig.\ref{edge_state}(c).
The edge states are predominantly chiral in nature, 
and the spin current at the two edges are opposite in direction (though
not equal in magnitude due to $D'\neq 0$, which breaks chiral symmetry).
For large $D'$  
($D'~\gg~D$), the tilting of the bands at the Dirac points 
is much greater and yields identical dispersion for the two edge states 
(Fig.~\ref{edge_state}(e)). This results in {\it anti}-chiral edge states
where the spin current is in the same direction along both edges of the
ribbon (Fig.~\ref{edge_state}(f)).
Finally, when $D~\approx~D'$, one of the edge
states (the top edge in the present case) acquires a dispersionless 
character (Fig.~\ref{edge_state}(h)). In other words, the edge state 
at the top is localized with no spinon transport while the bottom edge 
has a finite dispersion with a finite edge current 
(Fig.~\ref{edge_state}(i)).
Because of $U(1)$-symmetry of each spinon sector, there is a counter-propagating bulk current to compensate the imbalance between edge states.
However, the bulk current is not
uniform across the width of the ribbon. Instead, it is primarily confined 
to a small region near the edges.
At each edge, the bulk current opposes the edge current, with its 
magnitude decreasing rapidly away from the edges. 

We suggest that magnetic force microscopy(MFM) offers a promising experimental technique to measure the spinon current across the nano-ribbon and hence can detect the presence of anti-chiral edge states. Current MFM techniques can probe the local spin current in a finite sample to a resolution of a few nm. Since the topological character of the spinon bands for the different ranges of anisotropic DMI is reflected in distinct current profile across the ribbon, we believe MFM provides a promising experimental technique to identify anti-chiral edge states in real quasi-2D materials.
Additionally, inelastic neutron scattering spectra can also indirectly detect the presence of anti-chiral edge modes, by probing the magnon band structure. If the bands are tilted or the energy at $K$ and $K'$-point are unequal, it will suggest the presence of anti-chiral edge modes.

\begin{figure}[H]
	\centering
	\includegraphics[width=0.5\textwidth]{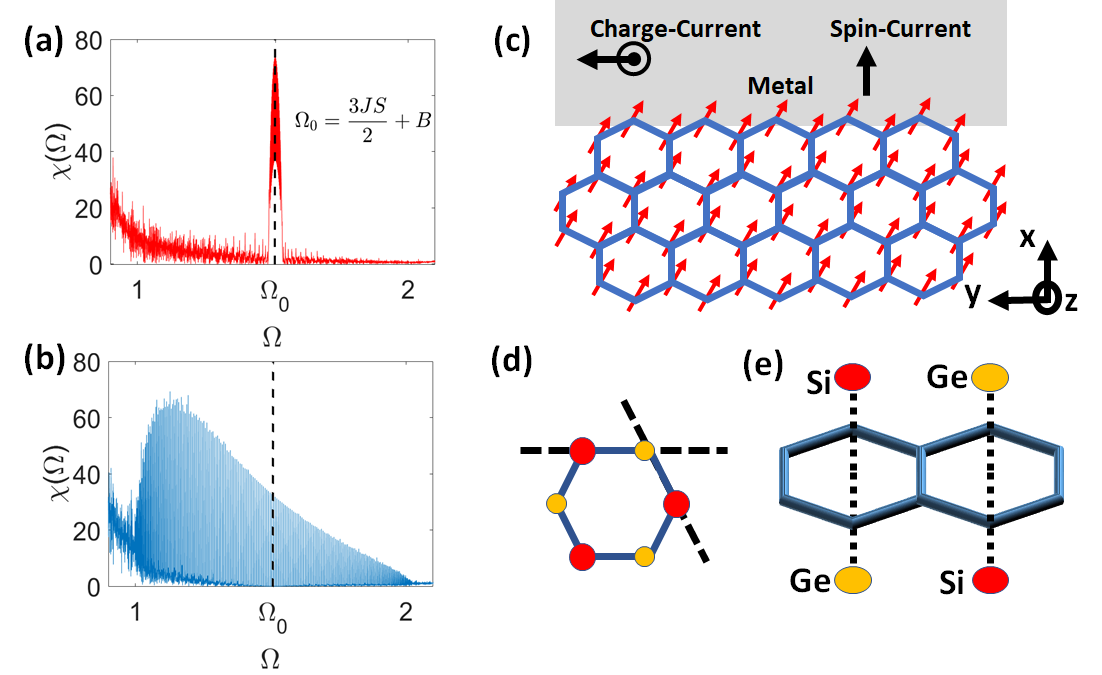}
	\caption{(color online) (color online) (a)-(b) Dynamical spin structure factor for two edges at temperature $T=0.4J$, with parameters $D=0.1, D'=0.09, B=0.01$. Red and blue colors denote upper and lower edges. (c) The experimental setup for spin Hall noise spectroscopy. (d) A ferromagnetic material with two different sub-latices with a mirror symmetric plane along the dashed lines. (e) A proposed material based on real materials \ce{CrGeTe3} and \ce{CrSiTe3} to realize antichiral edge states.}
	\label{ExperimentalRealization}
\end{figure}

Finally, we show that the dynamical-spin structure factor(DSSF) at the edge of the material\cite{SM} defined as  $
\chi(\Omega)=i\sum_{l\in \text{edge}}\left[ \chi_{ll}^{xx}(\Omega)+\chi_{ll}^{zz}(\Omega)\right],
$ where, $
\mathbin{\chi_{ij}^{\alpha\alpha}(\Omega)}{=-i\int^{\infty}_{-\infty} dt e^{-i\Omega t} \left\langle \hat{S}^\alpha_i(t) \hat{S}^\alpha_j(0) \right\rangle_0}
$
, offers a promising route to detecting anti-chiral edge states. The DSSF, shown
in the figure Fig.\ref{ExperimentalRealization}(a)-(b) for different edges, can be interpreted as the number of edge magnons present in a given energy level, and is proportional to the product of the density of states and Bose-Einstein distribution for the corresponding energy level. The signature of the edge states is
reflected in the features of the DSSF  near $\Omega=\frac{3J}{2}+B$, which is
the energy of the edge states in the absence of any DMI. Our results show that
$\chi(\Omega)$ is dramatically different for the two edges, when $\left|D\right|\approx \left|D'\right|$(equivalently $D_A\gg D_B$ or $D_B\gg D_A$). The DSSF can be measured using the experimental set-up shown in Fig.\ref{ExperimentalRealization}(c), according to reference\cite{SHNS}. The quantum fluctuation of the spins at the edges gives rise to spin current in metal and as a consequence the spin current gives rise to the charge current in transverse direction due to inverse spin Hall effect. Measurement of the noise-spectrum in the charge current gives the information of the DSSF.

The presence of anti-chiral DMI requires two in-equivalent sub-lattices in the 2D-honeycomb lattice, as shown in Fig.\ref{ExperimentalRealization}(d). 
The presence of two different types of atoms will result in asymmetric DMI,
leading to a broken inversion symmetry and non-zero $D'$. 
Mirror symmetry along the dotted lines prevents any non-zero perpendicular DMI on nearest-neighbour bonds, whereas in-plane mirror symmetry suppresses any in-plane DMI. While we
 \noindent  are not aware of any such material at present, the recent
discovery of ferromagnetic order in 2D limit of several 
\ce{Cr}-based compounds including \ce{CrI3}~\cite{CrI3}, \ce{CrBr3}~\cite{CrBr3,CrBr32,CrBr33}, \ce{CrSrTe3}\cite{CrSiTe3} and \ce{CrGeTe3}\cite{CrGeTe3} as well as the \ce{Fe}-based family of compounds
\ce{AFe_2(PO_4)_2  (A=Ba,Cs,K,La)}\cite{IronDirac}
offer great promise. These quasi-2D
materials consist of weakly Van Der Waals-coupled honeycomb ferromagets. Presence of chiral DMI in some members of this family~\cite{CrI3}has been established using inelastic neutron scattering spectroscopy. In materials like \ce{CrSrTe3} and \ce{CrGeTe3}, presence of inversion center at the center of honeycomb cell makes the two sub-lattices equivalent. The inversion symmetry can be removed by replacing every other \ce{Ge} atom by an \ce{Si} atom as depicted in Fig.\ref{ExperimentalRealization}(e). In a similar vein, replacement of \ce{P} atom by another Group V element in \ce{AFe_2(PO_4)_2  (A=Ba,Cs,K,La)}\cite{IronDirac} will break the inversion symmetry of the lattice. The breaking of inversion symmetry may, in principle, give rise to additional interactions in these materials, e.g.,
nearest neighbor DMI. 
However, we have verified that inclusion of additional interactions, including nearest neighbor DMI as well as 2nd and 3rd nearest neighbor Heisenberg interactions only modifies the linear dispersion of the edge states, and
does not suppress the appearance of anti-chiral edge states\cite{SM}.

In conclusion, we have studied a Heisenberg ferromagnet with additional
next nearest neighbor DMIs on a honeycomb lattice with broken
sublattice symmetry. The unequal DMI between atoms on different 
sublattices, together with the broken chiral symmetry results in the emergence 
of anti-chiral edge states, in addition to the normal chiral modes. This is 
manifested in unique spin current distribution across the width of a finite system
with ribbon geometry. We propose experimental probes to detect the presence of
anti-chiral edge states as well as a potential material where such states may be
realized experimentally.

Financial support from the Ministry of Education, Singapore, in the form
of grant MOE2016-T2-1-065 is gratefully acknowledged.

\onecolumngrid

\section{Schwinger Boson Mean Field Theory}
The parent spin Hamiltonian is, 
\begin{align}
\pazocal{H}=& -J\sum_{\left\langle i,j \right\rangle} \boldsymbol{S}_i\cdot\boldsymbol{S}_j+D\sum_{\left\langle\left\langle i,j \right\rangle\right\rangle} \nu_{ij} \hat{z}\cdot \left(\boldsymbol{S}_i\times \boldsymbol{S}_j\right) \nonumber \\
& +D'\sum_{\left\langle\left\langle i,j \right\rangle\right\rangle} \nu_{ij}' \hat{z}\cdot \left(\boldsymbol{S}_i\times \boldsymbol{S}_j\right)-B\sum_i S_i^z,
\label{eq3}
\end{align}
After implementing the Schwinger boson transformation and imposing the constraints, the spinon Hamiltonian reads,
\begin{align}
    \pazocal{H}_{sp}=&-\frac{J}{2}\sum_{\left\langle i,j\right\rangle}\left(\cdagU{i}\cdagD{j}\cndagD{i}\cndagU{j}+\cdagD{i}\cdagU{j}\cndagU{i}\cndagD{j}\right)\nonumber\\
    &-\frac{J}{4}\left(\cdagU{i}\cndagU{i}\cdagU{j}\cndagU{j}+\cdagD{i}\cndagD{i}\cdagD{j}\cndagD{j}+\cdagU{i}\cndagU{i}\cdagU{j}\cndagU{j}+\cdagD{i}\cndagD{i}\cdagD{j}\cndagD{j}\right)\nonumber\\
    &+\frac{iD}{2}\sum_{\left\langle\left\langle i,j \right\rangle\right\rangle} \nu_{ij} \left(\cdagU{i}\cndagD{i}\cdagD{j}\cndagU{j}-\cdagD{i}\cndagU{i}\cdagU{j}\cndagD{j}\right)\nonumber\\
    &+\frac{iD'}{2}\sum_{\left\langle\left\langle i,j \right\rangle\right\rangle} \nu'_{ij} \left(\cdagU{i}\cndagD{i}\cdagD{j}\cndagU{j}-\cdagD{i}\cndagU{i}\cdagU{j}\cndagD{j}\right)\nonumber\\
    &-\frac{B}{2}\sum_i\left(\cdagU{i}\cndagU{i}-\cdagD{i}\cndagD{i}\right)+\sum_i\lambda_i\left(\cdagU{i}\cndagU{i}+\cdagD{i}\cndagD{i}-2S\right)\nonumber\\
    &+3NJS^2-4SN\lambda
\end{align}

The bond operators have to be choosen such that the total number of spinon is conserved in the mean field Hamiltoian which is equivalent to $S_z$-conservation in terms of spin\cite{conservation}. Defining the bond-operators, $\CHI=\cdag{i}\cndag{j}$ and $\CHI =\left(\CHIU +\CHID\right)/2$, we can re-write the Hamiltonian as,
\begin{align}
    \pazocal{H}_{sp}=&-2J\sum_{\left\langle i,j\right\rangle} :\CHI^\dagger\CHI:\nonumber\\
    &-\frac{D}{2}\sum_{\left\langle\left\langle i,j\right\rangle\right\rangle} i\nu_{ij} \left(:\CHIU^\dagger\CHID:-:\CHID^\dagger\CHIU:\right)\nonumber \\
    &-\frac{D'}{2}\sum_{\left\langle\left\langle i,j\right\rangle\right\rangle} i\nu'_{ij} \left(:\CHIU^\dagger\CHID:-:\CHID^\dagger\CHIU:\right)\nonumber \\
    &-\frac{B}{2} \sum_{i} \left(\cdagU{i}\cndagU{i}-\cdagD{i}\cndagD{i}\right) \nonumber\\
    &+\lambda\sum_{is} \cdag{i}\cndag{i}-4SN\lambda+3NJS^2
\end{align}
New bond operators are defined so that corresponding mean field parameters are real:  $\hat{A}_{ij,s}=\frac{1}{2}\left(\hat{\chi}_{ij,s}+\hat{\chi}_{ji,s}\right)$, $\hat{B}_{ij,s}=\frac{\nu_{ij}}{2i}\left(\hat{\chi}_{ij,s}-\hat{\chi}_{ji,s}\right)$, $\hat{B'}_{ij,s}=\frac{\nu'_{ij}}{2i}\left(\hat{\chi}_{ij,s}-\hat{\chi}_{ji,s}\right)$.
Defining mean field parameters $\eta=\left\langle \CHI^\dagger \right\rangle=\left\langle \CHI \right\rangle$, $\zeta_s=\left\langle\hat{A}_{ij,s}^\dagger\right\rangle=\left\langle\hat{A}_{ij,s}\right\rangle$, $\xi_s=\left\langle\hat{B}_{ij,s}^\dagger\right\rangle=\left\langle\hat{B}_{ij,s}\right\rangle$,
$\xi'_s=\left\langle\hat{B'}_{ij,s}^\dagger\right\rangle=\left\langle\hat{B'}_{ij,s}\right\rangle$, the quartic terms of Hamiltonian can be decoupled into quadratic and the mean-field spinon Hamiltonian takes the form as,
\begin{align}
    \pazocal{H}_{sp}^{mf}
    =&-\eta J\sum_{\left\langle i,j \right\rangle,s} \left[ \cdag{i} \cndag{j}+\mathtt{H.c.} \right] \nonumber\\
&+\frac{D}{2}\sum_{\left\langle\left\langle ij \right\rangle\right\rangle, s}\left[ i \nu_{ij} s \zeta^{ch}_{-s} \cdag{i}\cndag{j}+\mathtt{H.c.}\right]\nonumber \\
&+\frac{D}{2}\sum_{\left\langle\left\langle ij \right\rangle\right\rangle, s} \left[ s\xi^{ch}_{-s} \cdag{i}\cndag{j} + \mathtt{H.c.} \right] \nonumber \\
&+\frac{D'}{2}\sum_{\left\langle\left\langle ij \right\rangle\right\rangle, s}\left[ i \nu'_{ij} s \zeta^{Ach}_{-s} \cdag{i}\cndag{j}+\mathtt{H.c.}\right]\nonumber \\
&+\frac{D'}{2}\sum_{\left\langle\left\langle ij \right\rangle\right\rangle, s} \left[ s\xi^{Ach}_{-s} \cdag{i}\cndag{j} + \mathtt{H.c.} \right] \nonumber \\
&+\sum_{i,s} \left(\lambda-\frac{sB}{2}\right) \cdag{i}\cndag{i}\nonumber\\
&+6NJ\eta^2-6DN\sum_s s\zeta_s\xi_{-s}-6D'N\sum_s s\zeta_s\xi'_{-s}-4SN\lambda+3NJS^2
\end{align}
Fourier transformation of the mean field Hamiltonian  to 
momentum-space yields,
\begin{equation}
\pazocal{H}_{sp}^{mf}=\sum_{\boldsymbol{k}\in \mathtt{B.Z.},s} \Psi^\dagger_{\boldsymbol{k},s} \left[g_s(\boldsymbol{k})I+\boldsymbol{h}_s(\boldsymbol{k})\cdot \boldsymbol{\sigma}\right] \Psi_{\boldsymbol{k},s}+E_0,
\label{eq7}
\end{equation} 
where, $\Psi_{\boldsymbol{k},s}^\dagger=\left(\hat{a}_{\boldsymbol{k},s}^\dagger,\hat{b}_{\boldsymbol{k},s}^\dagger\right)$. $\hat{a}_{\boldsymbol{k},s}^\dagger$ and $\hat{b}_{\boldsymbol{k},s}^\dagger$ are the creation operators for Schwinger bosons on sublattice-A and sublattice-B~(see Fig.1(a) of main text), respectively. $\boldsymbol{\sigma}_{\alpha}$($\alpha=x,y,z$) represents the Pauli matrices. The other terms are given by,
\begin{align}
g_s(\boldsymbol{k})=&-\frac{sB}{2}+\lambda+sD\xi_{-s} \gamma_c^\beta
+sD'\left(\xi'_{-s}  \gamma_c^\beta-\zeta_{-s}\gamma_s^\beta\right),\nonumber \\
h_s(\boldsymbol{k})=&\begin{pmatrix}
-J\eta\gamma_c^\alpha \\
J\eta\gamma_s^\alpha \\
-D s \zeta_{-s}^c\gamma_s^\beta
\end{pmatrix},\nonumber \\
E_0=& 6NJ\eta^2-6DN\sum_s s\zeta_s \xi_{-s}-6D' N\sum_s s\zeta_s \xi'_{-s}-4SN\lambda+3NJS^2,
\end{align}
where, $\gamma_c^\beta=\sum_{j} \cos(\boldsymbol{k}\cdot\boldsymbol{\beta}_j),\gamma_s^\beta=\sum_{j}\sin(\boldsymbol{k}\cdot\boldsymbol{\beta}_j),\gamma_c^\alpha=\sum_{j}\cos(\boldsymbol{k}\cdot\boldsymbol{\alpha}_j),\gamma_s^\alpha=\sum_j \sin(\boldsymbol{k}\cdot\boldsymbol{\alpha}_j)$ and the vectors $\boldsymbol{\beta}_j$ and $\boldsymbol{\alpha}_j$ are shown in figure Fig.1(a). $N$ is the number of unit cells in lattice. $E_0$ is the energy of the ground state and the energies of spinons are considered with respect to the ground state energy.

After diagonalizing the k-space Hamiltonian we get,
\begin{equation}
    \pazocal{H}_{sp}^{mf}=E_0+\sum_{\boldsymbol{k},s,\tau}E_s^\tau(\boldsymbol{k})\cdag{\boldsymbol{k},\tau}\cndag{\boldsymbol{k},\tau},
\end{equation}
where, the relative energies,
\begin{equation}
E_s^\tau(\boldsymbol{k})=g_s(\boldsymbol{k})+\tau \left|h_s(\boldsymbol{k})\right|,
\end{equation}
refer to the upper ($\tau=+1$) and the lower ($\tau=-1$) band for each spinon sectors $s=\pm 1$.

\begin{figure}[H]
	\centering
	\includegraphics[width=0.5\textwidth]{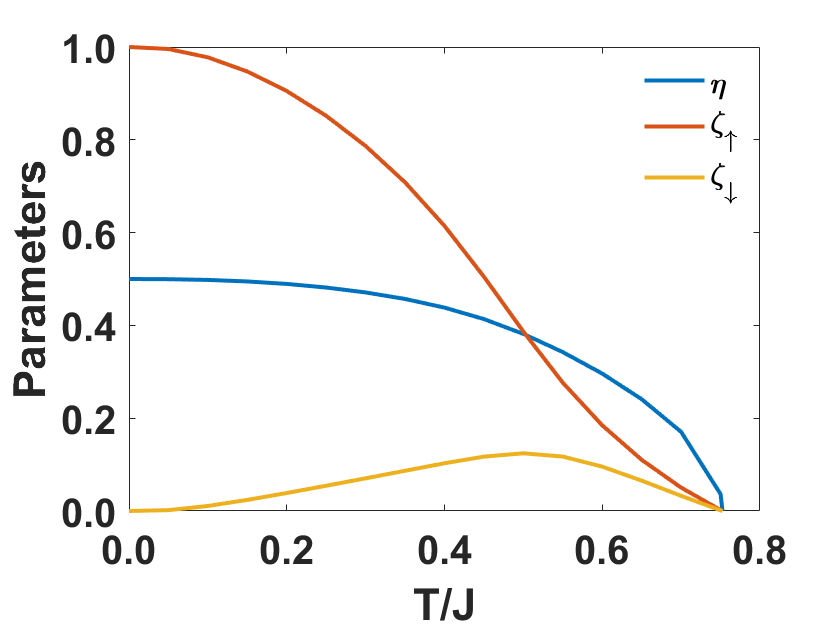}
	\caption{(color online)Plot of mean-field parameters, for $J=1.0, D=0.1, D'=0.05, B=0.1$}
	\label{parameter}
\end{figure}

From this we get the internal energy and the entropy of the non-interacting system as,
\begin{align}
    U&=E_0+\sum_{\boldsymbol{k},s,\tau} \rho_s^\tau(\boldsymbol{k})E_s^\tau(\boldsymbol{k}), \nonumber\\
    S&=k_B\sum_{\boldsymbol{k},s,\tau} [\left(1+\rho_s^\tau(\boldsymbol{k})\right) \ln\left(1+\rho_s^\tau(\boldsymbol{k})\right)-\rho_s^\tau(\boldsymbol{k})\ln \rho_s^\tau(\boldsymbol{k})],
\end{align}
where, $\rho_s^\tau(\boldsymbol{k})=\left[\exp\left(E^\tau_s(\boldsymbol{k})\right)-1\right]^{-1}$ 
is the Bose-Einstein distribution of spin-s spinons in the $\tau$-band. The Helmohltz-free-energy is given by,
\begin{align}
    G&=U-TS\nonumber\\
    &=E_0-{k_BT}\sum_{\boldsymbol{k},s,\tau}\ln\left(\frac{1}{1-\exp\left(\frac{-E_s^\tau(\boldsymbol{k})}{k_BT}\right)}\right)
\end{align}
After minimizing the Helmholtz free energy with respect to the mean field parameters, we get six self consistent equations, given by,
\begin{align}
2S=&\frac{1}{2N}\sum_{\boldsymbol{k},\tau,s} \rho_{s}^\tau(\boldsymbol{k}) \nonumber \\
1=-&\frac{J}{12N}\sum_{\boldsymbol{k},s,\tau} \tau \frac{\rho_s^\tau(\boldsymbol{k})}{\left| h_s \right|} \left|\sum_j e^{i\boldsymbol{k}\cdot\boldsymbol{\alpha}_j}\right|^2\nonumber \\
D\xi_s+D'\xi'_s = & \frac{1}{6N} \sum_{\boldsymbol{k},\tau} \left(D'- \tau s \frac{D^2 \zeta_{-s} \gamma}{h_s}\right)\rho_s^\tau(\boldsymbol{k}) \gamma_s^\beta\nonumber \\
\zeta_s=&\frac{1}{6N}\sum_{\boldsymbol{k},\tau} \rho^\tau_s(\boldsymbol{k}) \sum_j \cos(\boldsymbol{k}\cdot\boldsymbol{\beta}_j),
\end{align}
The mean field parameters are obtained by 
solving these six self-consistent equations. It is notable that the mean field parameter $\eta$ can be chosen as real, absorbing the complex phase factor into operator $\cndag{i}$. All other mean field parameters already chosen to be real. Using the parameters, we plot the band structure and evaluate corresponding topological information. For a fixed set of $J,D,D',B$, the mean field parameters are solved and plotted against temperature $T$ in Fig\ref{parameter}. The parameters 
$\eta$ and $\zeta_s$ represent short range correlations identifying magnetic
ordering and serve as order parameters for the ferromagnetic to paramegnetic
transition at higher temperatures\cite{correlation}. The constraint $\lambda$ is considered uniform throughout the lattice to retain the translational symmetry of the lattice.

 At low temperatures,
finite, non-zero values of $\eta$ and $\zeta_\uparrow$ denote ferromagnetic ordering. A positive $B$ determines that the spins are all aligned along the +ve
x-direction at $T=0$. In other words, the system is populated with up-spinons.
As the temperature increases, thermally excited down-spinons are generated, resulting
in a finite, non-zero $\zeta_\downarrow$. Finally at high temperatures, a vanishing of 
all the mean field parameters denote a transition to the paramagnetic phase. The paramagnetic phase transition with all zero correlations to be expected to be an outcome of large-N expansion. It has been shown for Heisenberg model that taking into account of the quantum fluctuations in the mean field parameter removes the phase transition\cite{mean_field_artifact}.

\section{Calculation of the edge-state and velocity distribution on a stripe geometry}
\label{sec2}
\begin{figure}[H]
	\centering
	\includegraphics[width=0.75\textwidth]{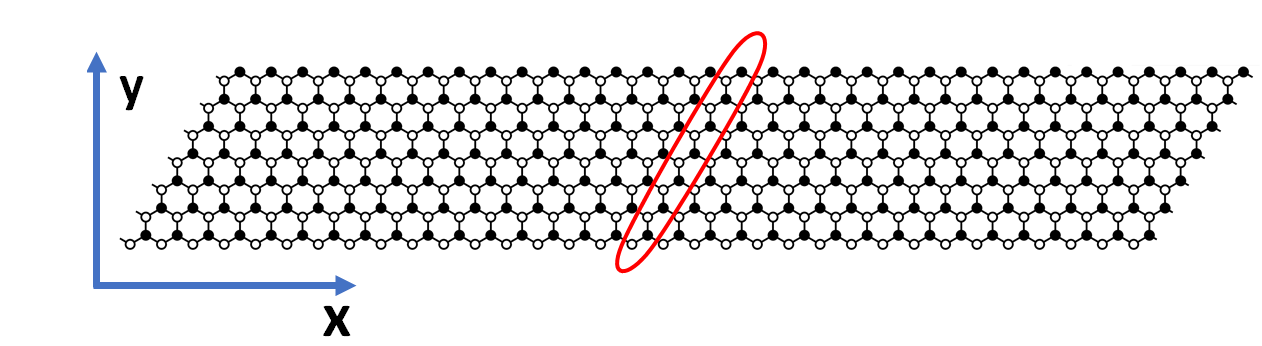}
	\caption{(color online)The honeycomb lattice structure of size ($20\times 14$) with periodic boundary condition along x-axis and open boundary condition along y-axis. The encircled sites are the basis of unit cell.}
	\label{stripe}
\end{figure}

The Hamiltonian in tight binding Hamiltonian can be written as,
\begin{equation}
    \pazocal{H}=\sum_{ij}t_{ij} \ket{i}\bra{j},
\end{equation}
where, i and j are the sites of lattice and $t_{ij}$ is the hopping amplitude. More explicitly the Hamiltonian can be written as a matrix form as\cite{book},
\begin{equation}
    \pazocal{H}=\sum_{m,s} \left[\Psi_{m,s}^\dagger U_s\Psi_{m,s}+\Psi_{m,s}^\dagger T_s\Psi_{m+1,s}+\Psi_{m+1,s}^\dagger T_s^\dagger\Psi_{m,s}\right],
    \label{MatrixForm}
\end{equation}
where, $\Psi_{m,s}^\dagger=(\hat{b}_{1,m,s}^\dagger,\hat{a}_{2,m,s}^\dagger,...,\hat{a}_{N,m,s}^\dagger,)$ and $N$ is the number of sites along the stripe(the sites inside the red circle of Fig.\ref{stripe} makes one stripe) and $m$ denotes the stripe index and $s$-denotes up or down spinon type. $U$ and $T$ are $N\times N$ matrices. Matrix-$U$ contains all the onsite and intra-stripe hopping elements(inside the red circle of Fig.\ref{stripe}) and matrix-$T$ contains all the inter-stripe hopping elements.

Imposing periodic boundary condition along the longitudinal direction as shown in the figure\ref{stripe}, one can Fourier transform the Hamiltonian with a 1D Bloch-wave vector, given by $\Psi_m=(1/\sqrt{M})\sum_{k=0}^{M-1} \Psi_k e^{-i2\pi km/M}$, where $M$ is the number of stripes along x-axis in Fig.\ref{stripe}. After Fourier transform the Hamiltonian can be written as,
\begin{align}
H &=\sum_k \Psi^\dagger_k \left[ U + \left(Te^{i\frac{2k\pi}{M}}+\text{H.c.}\right)\right]\Psi_k\nonumber\\
&=\sum_k \Psi^\dagger_k H_k \Psi_k,
\end{align}
The Hamiltonian can be diagonalized as,
\begin{equation}
\epsilon_k=P^\dagger H_k P,
\end{equation}
where, $P$ is an unitary matrix and the corresponding eigenvector is given by,
\begin{equation}
\Psi_k^d=P^\dagger\Psi_k
\label{eigenvector}
\end{equation}

 Diagonalizing the momentum space Hamiltonian, we obtain the bands for the stripe geometry, as shown in the figures Fig.3(b), 3(e), 3(h) of the main text.

The velocity operator, used to evaluate the spinon transport properties, are
expressed in terms of the k-space eigenstates as\cite{book2}\cite{discussion},
\begin{equation}
    \hat{v}=\sum_{ij}v_{ij} \ket{i}\bra{j},
    \label{VelocityOperator}
\end{equation}
where the coefficients $v_{ij}$ is given by,
\begin{equation}
    v_{ij}=\bra{i}\hat{v}\ket{j}=-\frac{i}{\hbar}\bra{i}[\hat{r},\pazocal{H}]\ket{j}=-\frac{i}{\hbar}(\boldsymbol{r}_i-\boldsymbol{r}_j)t_{ij}.
\end{equation}
We have calculated the distribution of the velocity component along x-axis across the cross section of the ribbon, which are shown in figure Fig.3(c),3(f),3(i) of the main text. 

\section{Calculation of the dynamical spin structure factor}
To calculate the dynamical spin structure factor, we have used the Holstein-Primakoff transformation, given by,
\begin{align}
\hat{S}^x_m=\frac{1}{2}\left(\hat{a}_m+\hat{a}_m^\dagger\right),\nonumber\\
\hat{S}^y_m=\frac{i}{2}\left(\hat{a}_m-\hat{a}_m^\dagger\right),\nonumber\\
\hat{S}^z_m=S-\hat{a}_m^\dagger\hat{a}_m,
\label{HP}
\end{align}
where, $\hat{a}_m^\dagger$ and $\hat{a}_m$ are the creation annihilation operators of Holstein-Primakoff bosons. 
For the case of ferromagnet with up spin at each site, the Holstein-Primakoff boson represents the down-spinons in Schwinger boson picture at low temperature. The diagonalized Hamiltonian can be written as,
\begin{equation}
H_k=\sum_n\epsilon_n(k) \hat{a}_{n,k}^{d\dagger}\hat{a}_{n,k}^{d},
\end{equation}
where, $\hat{a}^d_{n,k}$ is the bosonic operator after diagonalization. Using the above relation and Heisenberg's equation of motion it can be proved that,
\begin{align}
\hat{a}_{n,k}^{d\dagger}(t)&=\hat{a}_{n,k}^{d\dagger}(0)e^{i\epsilon_n(k)t}\nonumber\\
\hat{a}_{n,k}^{d}(t)&=\hat{a}_{n,k}^{d}(0)e^{-i\epsilon_n(k)t}
\label{Heisenberg}
\end{align}

 The dynamical spin structure factor in terms of Holstein-Primakoff Boson, is given by,
\begin{align}
\chi(\Omega)&=\sum_{m\in \text{edge}} \left[\chi^{xx}_{mm}(\Omega)+\chi^{zz}_{mm}(\Omega)\right]\nonumber\\
&=\sum_{m\in \text{edge}} \left[\int^\infty_{-\infty} dt e^{-i\Omega t}\left(-i\left\langle \hat{S}^x_{1,m}(t) \hat{S}^x_{1,m}(0)\right\rangle\right)+\int^\infty_{-\infty} dt e^{-i\Omega t}\left(-i\left\langle \hat{S}^z_{1,m}(t) \hat{S}^z_{1,m}(0)\right\rangle\right)\right]\nonumber\\
&=-i \sum_m \int^{\infty}_{-\infty} dt e^{-i\Omega t} \left[ \frac{S}{2}\left\langle \hat{a}_{1,m}(t)\hat{a}_{1,m}(0) + \hat{a}_{1,m}(t)\hat{a}_{1,m}^\dagger(0) + \hat{a}_{1,m}^\dagger(t)\hat{a}_{1,m}(0) + \hat{a}_{1,m}^\dagger(t)\hat{a}_{1,m}(0)\right\rangle\right.\nonumber\\
&\qquad \qquad\qquad  \left. +\left\langle S^2-S\hat{a}_{1,m}^\dagger (0)\hat{a}_{1,m}(0)-S\hat{a}^\dagger_{1,m}(t)\hat{a}_{1,m}(t)\right\rangle\right] \qquad\qquad\text{[Neglecting the higer order terms]}.
\label{chi}
\end{align}
Transforming the boson operator $\hat{a}_{1,m}(t)$ into diagonalized boson operator $\hat{a}^d_{1,m}(t)$ and
 using the relation Eq.\ref{Heisenberg}, we derive the spin-structure factor.
 \raggedbottom
 \pagebreak
\section{Modulation of edge state dispersion in presence of other interactions in a honeycomb ferromagnet}
\begin{figure}[H]
	\centering
	\includegraphics[width=0.3\textwidth]{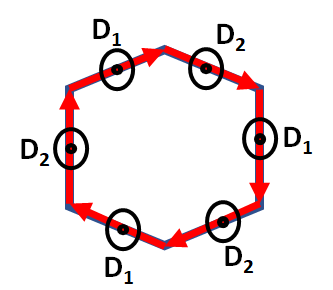}
	\caption{(color online) DM-interactions on nearest neighbour bonds.}
	\label{DMinteraction}
\end{figure}
The spin Hamiltonian studied in theis work is an idealized model. For example, the long range Heisenberg interactions are neglected. Moreover, the breaking of inversion symmetry required for anti-chiral DM-term $D'$ might give rise the nearest neighbour out of plane DM-interactions and also in plane DM-interactions. At low temperature, the in plane interactions can be neglected, which gives rise to three magnon interactions in terms of Holstein Primakoff Bosons. 
Neglecting, any presence of in plane DM-interaction at low temperature, we can re-write a more general Hamiltonian of the material as,
\begin{align}
\pazocal{H}=J_1 \sum_{\left\langle i,j \right\rangle} \boldsymbol{S}_i &\cdot \boldsymbol{S}_j+
 J_2 \sum_{\left\langle\left\langle i,j \right\rangle\right\rangle} \boldsymbol{S}_i \cdot \boldsymbol{S}_j+
 J_3 \sum_{\left\langle\left\langle\left\langle i,j \right\rangle\right\rangle\right\rangle} \boldsymbol{S}_i \cdot \boldsymbol{S}_j\nonumber\\
 &+D_1\sum_{\left\langle ij\right\rangle _A} \nu_{ij}\hat{z}\cdot(\boldsymbol{S}_i \times \boldsymbol{S}_j)+
 D_2\sum_{\left\langle ij\right\rangle_B} \nu_{ij}\hat{z}\cdot(\boldsymbol{S}_i \times \boldsymbol{S}_j)\nonumber\\
 &+D\sum_{\left\langle\left\langle ij\right\rangle\right\rangle_B} \nu_{ij}\hat{z}\cdot(\boldsymbol{S}_i \times \boldsymbol{S}_j)+D'\sum_{\left\langle\left\langle ij\right\rangle\right\rangle_B} \nu'_{ij}\hat{z}\cdot(\boldsymbol{S}_i \times \boldsymbol{S}_j)\nonumber\\
 &+B\sum_i \boldsymbol{S}_i,
 \label{Ham}
\end{align}
where, the DM-interactions $D_1$ and $D_2$ are defined on the nearest neighbour bonds $\left\langle ij\right\rangle _A$ and $\left\langle ij\right\rangle _B$ as shown in Fig.\ref{DMinteraction}. Any single ion anisotropy terms acts as chemical potential for the spin-excitation, and is accounted for by a renormalization of the magnetic field. As a more realistic model, we have fixed the Heisenberg interactions present in the material \ce{CrI3}\cite{CrI3}, $J_1=2.09$ meV, $J_2=0.16$ meV, $J_3=0.18$ meV. The nearest neighbour DM-terms and magentic field are fixed as $D_1=0.1$ meV, $D_2=0.15$ meV, $B=g\mu_B B_z=0.01$ meV. We transformed the spin Hamiltonian into magnon Hamiltonian using Holstein Primakoff transformation defined in Eq.\ref{HP}. Then, plotted the band structure and dynamical spin structure factor in Fig\ref{Modified}. Our results confirm that the qualitative behaviour is same as the ideal model considered in the test and the behaviour of edge states mostly depends on the DM-interactions $D$ and $D'$. The presence of other interaction terms in the Hamiltonian just distorts the linear dispersion of the edge states to a non-linear dispersion.

\begin{figure}[H]
	\centering
	\includegraphics[width=1\textwidth]{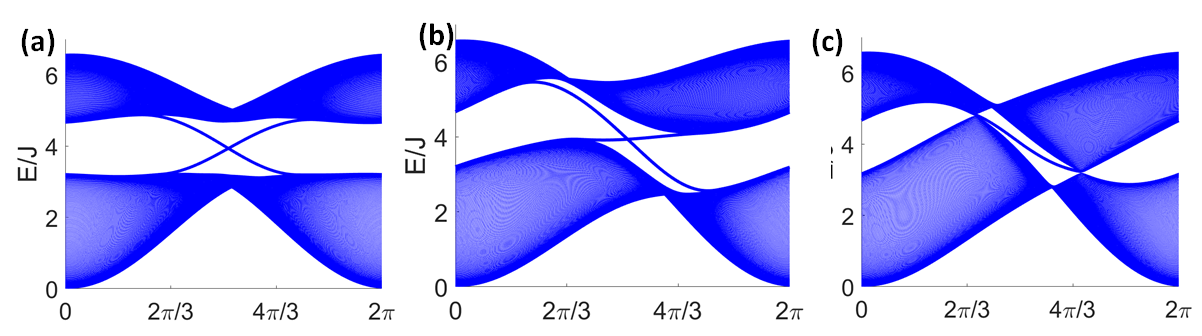}
\end{figure}
\begin{figure}[H]
	\centering
	\includegraphics[width=1\textwidth]{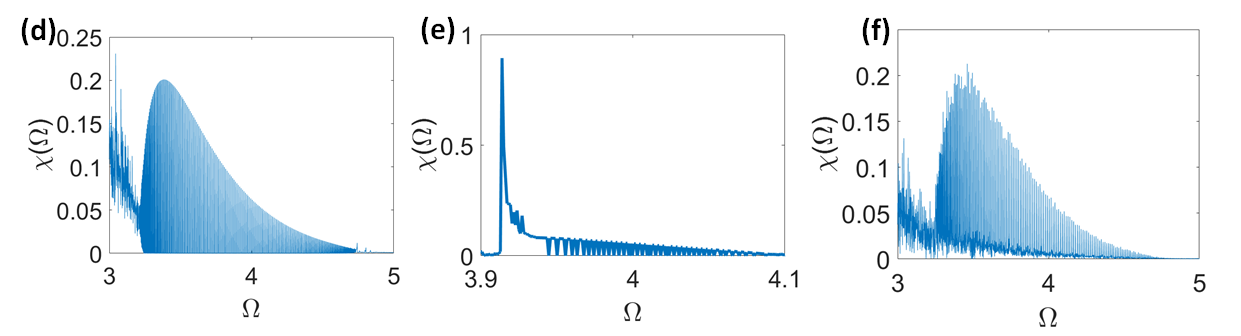}
\end{figure}
\begin{figure}[H]
	\centering
	\includegraphics[width=1\textwidth]{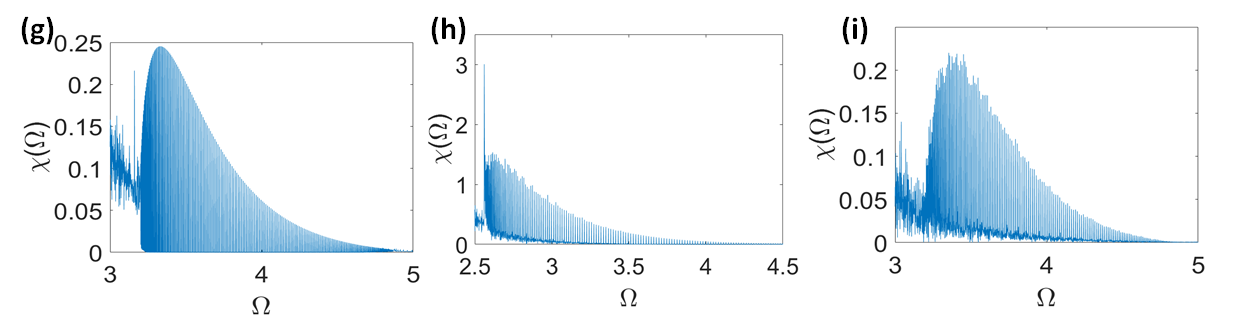}
	\caption{(color online) Band structure of magnons of a stripe geometry for parameters (a) $D=0.31 meV$, $D'=0.01 meV$, (b) $D=0.31 meV$, $D'=0.28 meV$, (c) $D=0.01 meV$, $D'=0.31 meV$. The dynamical spin structure factor for upper edge at $T=0.4$ for parameters, (d) $D=0.31 meV$, $D'=0.01 meV$, (e) $D=0.31 meV$, $D'=0.28 meV$, (f) $D=0.01 meV$, $D'=0.38 meV$. The dynamical spin structure factor for lower edge at $T=0.4$ for parameters, (g) $D=0.31 meV$, $D'=0.01 meV$, (h) $D=0.31 meV$, $D'=0.28 meV$, (i) $D=0.01 meV$, $D'=0.38 meV$.}
	\label{Modified}
\end{figure}

\bibliographystyle{apsrev4-1}

\end{document}